\documentclass[%
 reprint,
superscriptaddress,
nofootinbib,
amsmath,amssymb,
aps,
11pt,
onecolumn
]{revtex4-1}
\usepackage[utf8]{inputenc}
\usepackage{multirow}
\usepackage{graphicx}
\usepackage{dcolumn}
\usepackage{booktabs}
\usepackage{makecell}
\usepackage{threeparttable}
\usepackage{bm}

\usepackage{siunitx}
\usepackage{hyperref}
\usepackage{footnote}
\usepackage{tikz}
\usepackage{tikz-feynman}
\usepackage{float}
\allowdisplaybreaks[4]

\newcommand{\bracket}[2]{\langle #1 |#2\rangle}

\newcommand{\itp}{\affiliation{CAS Key Laboratory of Theoretical Physics, Institute of Theoretical Physics,\\
Chinese Academy of Sciences, Beijing 100190, China}}
\newcommand{\ucas}{\affiliation{School of Physical Sciences, University of Chinese Academy of Sciences, Beijing 100049, China}}
\newcommand{\ihep}{\affiliation{Institute of High Energy Physics, Chinese Academy of Sciences, Beijing, 100049, China}}

\begin{abstract}
    In this paper, we propose a method to construct the decay amplitudes in the orbital ($L$) and spin ($S$) coupling scheme for particles with arbitrary spins. For the $1\to 2$ decay with only massive particles involved, the angular dependence is completely encoded in the angular momentum part, and the spins of daughter particles are coupled in the rest frame of the mother particle, which contributes only a constant factor. For the sequential decay, the total amplitude is constructed by the two $1\to2$ amplitudes evaluated in the rest frame of their own mother particles, and then they are transformed to the common frame, usually chosen as the laboratory frame, by certain Lorentz transformations. In this way, it is easy to add the amplitudes of possible different decay chains coherently. If massless particles show up in the final states, the polarizations are expressed in helicity basis and the amplitudes are modified correspondingly.
\end{abstract}

\begin{document}
\title{Spin-orbit amplitudes for decays with arbitrary spin}
\author{Xiao-Yu Li}\email{xiaoyuli@ihep.ac.cn}
\ihep\ucas
\author{Xiang-Kun Dong}\email{dongxiangkun@itp.ac.cn}
\itp\ucas
\author{Hao-Jie Jing}\email{jinghaojie@ucas.ac.cn}
\ucas



\maketitle

\section{Introduction}
The underlying theory describing the strong interaction, the quantum chromodynamics (QCD), is nonperturbative at low energy, and thus it still remains a challenge to describe the low energy interaction between hadrons from the first principle. On the other hand, the high quality and a vast amount of data on hadron productions and decays are or will be available in the worldwide high energy experiments. Hence rigorous amplitude analysis is necessary to extract the parameters of resonances. In some cases, the resonances have apparent signals in the data, and simple fit models, e.g., Breit-Wigner or Flatt\`e,  work well to extract the resonance parameters. However, there are many other cases where several resonances live in the same energy region and overlap. In such cases, the partial wave analysis is necessary. 

Several methods have been used to construct amplitudes that are used to fit the experimental data. By taking the hadrons as elementary particles and writing down the effective Lagrangian, one can obtain the corresponding amplitudes, see, e.g., Refs~\cite{Olsson:1974cn,Benmerrouche:1994uc,Nakayama:2002mu,Liang:2002tk}. This method keeps the Lorentz invariance, but it is hard to obtain individual amplitudes with certain orbital angular momentum. In the helicity-coupling amplitudes proposed in Ref~\cite{Jacob:1959at} and developed in Refs~\cite{Chung:1971ri,Chung:1993da,Chung:1997jn}, the amplitudes are constructed to represent the processes where final states are in the helicity basis, and the angular dependence is encoded in the Wigner $D$ matrix for rotations. The helicity-coupling amplitudes for sequential decays with several decay chains have been derived in Ref~\cite{Chen:2017gtx,Marangotto:2019ucc} so that they can be added coherently. Another way to write the amplitudes for a certain spin and orbital angular momentum is called covariant $L$-$S$ scheme~\cite{Zou:2002ar,Zou:2002yy}\footnote{It is called the spin-momentum operator expansion in Refs.\cite{Anisovich:2004zz,Anisovich:2006bc}.} where the blocks are the spin wave functions of initial and final particles and the tensors for orbital angular momenta. In this scheme, the independent amplitudes for a certain decay are characterized by the total spin of final states and the orbital angular momentum. This method holds the Lorentz covariance of each block, and different decay chains can be added coherently. However, it is very {complicated} to construct the total spin wave functions if high spin particles are involved. In Ref.~\cite{JPAC:2019ufm}, the Dalitz-plot
decomposition is proposed to deal with the three-body decays. The comparisons between some of these methods can be consulted in, e.g., Refs~\cite{JPAC:2017vtd,JPAC:2018dfc}

In this paper, we overcome the drawbacks of the covariant $L$-$S$ coupling method and provide a systematic way to construct the decay amplitudes with certain orbital angular momentum and total spin of the final particles. Unlike the treatments in Refs~\cite{Zou:2002ar,Zou:2002yy} where explicit couplings of spin wavefunctions are needed{, which are usually complicated when high spin particles are involved}, we introduce an abstract coupling function that couples two spins into one total spin. Besides, the spins are defined in their ``standard" frame, and the angular dependence is totally absorbed into the angular momentum tensors for massive particles, contrary to those in Refs~\cite{Zou:2002ar,Zou:2002yy} where the spin wave functions still depend on angles.

This paper is organized as follows: The spin wave functions for arbitrary spins in arbitrary frames are introduced in Sect.~\ref{sec:II}; The covariant $L$-$S$ coupling amplitudes for $1\to2$ decays are constructed in Sect.~\ref{sec:III} and the sequential decays are considered in Sect.~\ref{sec:IV}, followed by several illustrative examples presented in Sect.~\ref{sec:V}; We finally give the discussion in Sect.~\ref{sec:VI}

\section{Spin wave functions}\label{sec:II}
 In this work, we define the spin wave function of a particle in its standard frame (STDF). A STDF is a reference frame where the particle in question has a standard momentum (STDM)~\cite{Weinberg:1995mt}, namely,
\begin{align}
    k^\mu=\left\{\begin{array}{ll}
       (M,0,0,0)  & \text{for mass $M$} \\
       (\kappa,0,0,\kappa) & \text{for mass $0$}
    \end{array}\right.,
\end{align}
with $\kappa$ a positive energy. Associating with the STDM there arises the corresponding little group (CLG). Under a transformation in CLG the STDM is invariant. For example, the STDM of a massive particle $(M,0,0,0)$ is invariant under a SO(3) rotation transformation. It could be seen that for a given particle there can be multiple choices of STDF. These STDFs are connected by CLG transformation. The choice of STDF for a given particle to define its polarization is just a matter of convention. 

    \begin{table}
    \begin{tabular}{l|c|c}
    \hline
     & Massive Particle & Massless Particle \\
    \hline
    STanDard Momentum(STDM) & $(0,0,0,M)$ & $(\kappa,0,0,\kappa) $\\
    Corresponding Little Group(CLG) & SO(3) & ISO(2) \\
    Polarization Coefficients between STDFs & Wigner-$D$ Matrix & Gauge Phase \\
    STanDard Boost(STDB) & $RBR^{-1}$ & $RB$ \\
    \hline
    \end{tabular}
    \end{table}

In the STDF, the spin wave function reads
\begin{align}
   \chi_S^{\mu_1\cdots\mu_n}(m)=\left\{ \begin{array}{ll}
        \epsilon^{\mu_1\cdots\mu_n}(m) & \text{for a boson with }S=n \\
        u/v^{\mu_1\cdots\mu_n}(m) & \text{for a fermion/antifermion with }S=n+1/2
    \end{array}\right.,\label{eq:chiSp}
\end{align}
with $m$ the polarization. For a massive particle with spin-1, we have
\begin{align}
    \chi_1^\mu(0)=(0,0,0,1)\quad\chi_1^{\mu}(\pm1)=\frac{1}{\sqrt2}(0,\mp1,-i,0)\label{eq:spin1},
\end{align}
while for spin-1/2, they take the following form in the Weyl basis
\begin{align}
    u_{1/2}(1/2)=\sqrt{M}(1,0,1,0)^{\rm T},\quad  u_{1/2}(-1/2)=\sqrt{M}(0,1,0,1)^{\rm T},\notag\\
    v_{1/2}(1/2)=\sqrt{M}(0,-1,0,1)^{\rm T},\quad  v_{1/2}(-1/2)=\sqrt{M}(1,0,-1,0)^{\rm T},
\end{align}
For a massless spin-1 particle, say photon, the spin wave function takes the same form as those in Eq.(\ref{eq:spin1}) except that $m=0$ is not allowed. The wave functions for particles with higher spins in their ``standard" frame can be constructed via $ \chi_1^\mu(m)$ and $ u/v_{1/2}(m)$ together with Clebsch-Gordan (CG) coefficients~\cite{Rarita:1941mf,Zou:2002yy}. 

Polarizations of a given particle defined in two STDFs are connected by a linear transformation, this is well known as a Wigner-rotation in massive particle case. From this we see that different choice of STDF just means a change of polarization basis.

The spin wave function of a particle with arbitrary momentum $p$ is defined by
\begin{align}
    \chi_S(p,m)\equiv U(\tilde\Lambda_{k\to p}) \chi_S(m),\label{eq:chidef}
\end{align}
where $\tilde\Lambda_{k\to p}$ is a Lorentz transformation that takes $k$ to $p$ and $U(\Lambda)$ is the corresponding representation matrix acting on $\chi_S$, and we have dropped the Lorentz indices for simplicity.
Note that all the Lorentz transformations $\Lambda_{k\to p}$ that take $k$ to $p$ form a coset of the CLG $H_k$ for the ``standard" momentum $k$ (SO(3) for massive particles and ISO(2) for the massless ones). $\tilde\Lambda_{k\to p}$ belongs to this coset and it is convenient to choose the standard boost (STDB)
\begin{align}
    \tilde\Lambda_{k\to p}=\left\{
    \begin{array}{ll}
       R(\hat{\bm p})B\left(\sqrt{\frac{\bm p^2}{M^2+\bm p^2}}\right)R^{-1}(\hat{\bm p})  & \text{ for mass $M$,} \\
        R(\hat{\bm p})B\left(\frac{\bm p^2-\kappa^2}{\bm p^2+\kappa^2}\right) & \text{ for mass 0,}
    \end{array}\right.\label{eq:standboost}
\end{align}
where $R(\hat{\bm p})$ rotates $z$ axis to the direction $\hat{\bm p}$ and $B(v)$ is the boost along $z$-axis by velocity $v$.\footnote{ Using such STDB, the spin wave function takes the canonical form for a massive particle while takes the helicity form for a massless particle} A general Lorentz transformation can be decomposed as
\begin{align}
    \Lambda_{k\to p}=\tilde\Lambda_{k\to p}\Lambda_{k\to k},
\end{align}
where $\Lambda_{k\to k}$ belongs to $H_k$. Under this Lorentz transformation, the spin state is transformed as
\begin{align}
     \chi_S(p,m)&=\sum_{m'}D^S_{mm'}(\Lambda_{k\to k})U(\tilde\Lambda_{k\to p})\chi_S(m')\notag\\
     &=\sum_{m'}D^S_{mm'}(\Lambda_{k\to k})\chi_S(p,m'),
\end{align}
with $D^S_{mm'}$ the representation matrix for $H_k$\cite{Weinberg:1995mt}.


\section{$L-S$ coupling for two-body decay processes}\label{sec:III}
Consider a two-body decay process
\begin{align}
a^{S_a}(p_a,m_a)\to b^{S_b}(p_b,m_b)\overbrace{+}^{L,S}c^{S_c}(p_c,m_c),
\end{align}
where $S_i,m_i$ and $p_i$ are the spin, polarization and momentum of particle $i$, respectively, defined in an arbitrary frame, say the laboratory frame. $L$ is the orbital angular momentum between the final two particles in this frame, and $S$ is the total spin of the final state. Note that $m_i$ is the same as that in the STDF of particle $i$. It is convenient to construct the amplitude in the STDF of the mother particle $a$, denoted by $a^*$-frame, which is obtained by the Lorentz transformation $\tilde \Lambda_{p_a\to k_a}$ from the laboratory frame. In $a^*$-frame, the momenta and polarizations are labeled by $p_i'$ and $m_i'$, respectively, with $p^{\nu}_i=\tilde\Lambda^{\mu\nu}_{k_a\to p_a}p_{i\mu}'$, $p_a'=k_a$ and
\begin{align}
    \chi_{S}(m_i)=\sum_{m_i'}D^S_{m_i,m_i'}(\Lambda^{a}_{k_i\to k_i})\chi_S(m_i'),
\end{align}
with $\Lambda^{a}_{k_i\to k_i}=\tilde\Lambda^{-1}_{k_i\to p_i}\tilde\Lambda_{k_a\to p_a}\tilde\Lambda_{k_i\to p'_i}$.
The spatial wave function of particle $b$ and $c$ can be decomposed into angular and radial parts and the former, which can couple to the spins, will be represented by the orbital angular momentum tensor introduced in Refs~\cite{Chung:1993da,Chung:1997jn},
\begin{align}
    T_L^{\mu_1\cdots\mu_L}(p'_b,p'_c,m_L')={q'^{\nu_1}\cdots q'^{\nu_L}\epsilon^*_{\nu_1\cdots\nu_L}(m_L')}\epsilon_L^{\mu_1\cdots\mu_L}(m_L')B_L(Q_{abc}),\label{eq:L}
\end{align}
with $q'=p'_b-p'_c$ and $m_L'$ the $z$-component of orbital angular momentum, and $\epsilon_L^{\mu_1\cdots\mu_L}$ is constructed following Refs~\cite{Rarita:1941mf,Zou:2002yy}. We have introduced a Blatt-Weisskopf barrier factor $B_L$ as in Refs~\cite{VonHippel:1972fg,Chung:1993da,Zou:2002yy,Zou:2002ar} to suppress the high energy tail of the orbital angular momentum tensor with $Q_{abc}$ the magnitude of the 3-momentum of particle $b$ in the $a^*$-frame. Note that the Blatt-Weisskopf barrier factor is model dependent and one can introduce other forms of the suppression factor.

In general, the spin wave functions of a particle depends on its momentum and in turn introduces additional angular dependence to the amplitude, as defined in Eq.(\ref{eq:chidef}). To absorb such angular dependence into the orbital momentum, one can express the $\chi_S(p,m)$ as that in a STDF transformed by a Lorentz transformation, see Eq.(\ref{eq:chidef}). When constructing the amplitudes, these Lorentz transformation can be absorbed into the angular momentum tensors defined in Eq.(\ref{eq:L}) and the spin part constructed with the spin wavefunctions with standard momenta are now independent of angle, see Ref.~\cite{Jing:2023rwz} for more explanations.

The general expression of the amplitude $\mathcal M$ can be constructed as a Lorentz and dirac scalar function of the momentum $p_i$ and spin wave functions and can be decomposed into different partial-wave contributions,
\begin{align}
\mathcal M_{a\to bc}(m'_a;p'_b,m'_b,p'_c,m'_c;m'_L) = \sum_{L,S}  A_{L,S}^{S_a} \mathcal M^{S_a}_{L,S},
\end{align}
where $ A^{S_a}_{LS}$ are unknown, usually complex, dynamic parameters related to certain spatial angular momentum and total spin and 
$\mathcal M^{S_a}_{LS}$ is the partial component with certain $L$ and $S$, which can be expressed as
\begin{equation}
\mathcal M^{S_a}_{L,S}\left( m'_a;p'_{b},m'_b,p'_{c},m'_c;m'_L \right) = J^{S_a;L,S*}_{\mu_{1}\cdots\mu_n} \left( m'_a;p'_{b},m'_b,p'_{c},m'_c;m'_L \right){\chi}_{S_a}^{\mu_{1}\cdots\mu_n}(m'_a),
\label{eq:MLS}
\end{equation}
where $J^{S_a;L,S}$ denotes the total angular momentum ($S_a$) wave function of the final states and it should be constructed by wave functions of $L$, $S_b$, and $S_c$, as given below in Eq.(\ref{eq:LS}).

In order to construct $J^{S_a;L,S}$, we introduce the coupling function, which is linear on its arguments and defined by
\begin{align}
    F^{[S,m]}\left(\chi^{\ }_{S_1}(m_1),\chi^{\ }_{S_2}(m_2)\right)=C_{S_1 m_1 S_2 m_2}^{S m} \chi_S(m),
\end{align}
with $C^{jm}_{j_1m_1j_2m_2}\equiv\bracket{j_1m_1;j_2m_2}{jm}$ the CG coefficients of SU$(2)$, to compose and project wave functions of two massive particles into a combined wave function. 
It is easy to realize that this coupling function is similar to $C^{jm}_{j_1m_1j_2m_2}$, which combines two states with definite spin $j_1$ and $j_2$ into a state with spin $j$; here the coupling function $F^{[S,m]}\left(\chi^{\ }_{S_1}(m_1),\chi^{\ }_{S_2}(m_2)\right)$ combines two Lorentz covariant wave functions $\chi^{\ }_{S_1}(m_1)$ and $\chi^{\ }_{S_2}(m_2)$ into another Lorentz covariant wave function $\chi_S(m)$. Therefore, such coupling functions correspond to the Lorentz covariant CG coefficients of the group SU(2). Through group theory, especially by using irreducible tensors, one can find the explicit forms of these coupling functions, so as to construct the $L$-$S$ coupling satisfying the Lorentz covariance. The $F$ function is actually a projection operator, which performs direct-product and decomposition to the two arguments, and projects the result into a coupled spin wave function. The component 
form of a $F$ function, denoted as $\left[F^{[S,m]}\left(\chi^{\ }_{S_1}(m_1),\chi^{\ }_{S_2}(m_2)\right)\right]_{\rho_1\cdots\rho_a\sigma_1\cdots\sigma_b}$, is 
\begin{align}
\chi_{S,\rho_1\cdots\rho_a\sigma_1\cdots\sigma_b}(m)
{G}^{\mu_1 ... \mu_a,\nu_1 ... \nu_b}_{S,S_1,S_2}(m)\chi_{S_1,\mu_1 ... \mu_a}(m_1)\chi_{S_2,\nu_1 ... \nu_b}(m_2),
\end{align}
with 
\begin{align}
G^{\mu_1 ... \mu_a,\nu_1 ... \nu_b}_{S,S_1,S_2}(m)=\sum_{m'_1,m'_2}C^{Sm}_{S_1m'_1S_2m'_2}{\chi}^{\mu_1 ... \mu_a*}_{S_1}(m'_1){\chi}^{\nu_1 ... \nu_b*}_{S_2}(m'_2).
\end{align}

Similarly, the orbital angular momentum tensor is from a projection which projects coupling irreducible tensors and representation matrices of STDB of the two daughter particles into a spin wave function standing for spatial part. The strict mathematical basis of this function is discussed in Ref.~\cite{Jing:2023rwz} and here just keep in mind that it does yield the correct coupling. 

When massless particles are involved, we need an additional rotation to transform the STDF of the massless particle to the frame that has the same $z$-direction as that of the mother particle. Therefore, the arguments of the coupling function $F$ are the polarization functions after such rotation for massless particles. To be specific, in a two-body decay process if one of the final particles, say particle 2, is massless, the coupling reads
\begin{align}
    F^{[S,m]}\left(\chi^{\ }_{S_1}(m_1),\tilde\chi^{\ }_{S_2}(m_2')\right)=\sum_{m_2'}D^{S_2}_{m_2,m_2'}(R(\hat{\bm p}_{2}))C_{S_1 m_1 S_2 m_2'}^{S m} \chi_S(m),\label{eq:F1}
\end{align}
while if two particles are both massless, the coupling should be 
\begin{align}
    F^{[S,m]}\left({\tilde\chi^{\ }_{S_1}(m_1')},\tilde\chi^{\ }_{S_2}(m_2')\right)=\sum_{m_1',m_2'}D^{S_1}_{m_1,m_1'}(R(\hat{\bm p}_{1}))D^{S_2}_{m_2,m_2'}(R(\hat{\bm p}_{2}))C_{S_1 m_1' S_2 m_2'}^{S m} \chi_S(m),\label{eq:F12}
\end{align}
 where 
 \begin{align}
 \tilde\chi^{\ }_{S_i}(m_i)=\sum_{m_i'}D^{S_i}_{m_i,m_i'}(R(\hat{\bm p}_{i}))\chi^{\ }_{S_i}(m_i')
 \end{align}
 is the polarization of particle $i$ after rotation, and $R(\hat{\bm p}_{i})$ rotates $z$-axis to the direction of the three-momentum of particle $i$. In the following, we will focus on the cases where all particles are massive and generalization to the massless case is straightforward by using Eqs.(\ref{eq:F1},\ref{eq:F12}).

In terms of the coupling function, the total spin wave function of the final two particles reads
\begin{align}
    \chi_{S}^{\rm tot}(m'_S)=F^{[S,m'_S]}\left(\chi_{S_b}(m'_b),\chi_{S_c}(m'_c)\right),\label{eq:chis}
\end{align}
where $m'_S=m'_b+m'_c$. 
 Similarly, the total wave function of the final states in Eq.(\ref{eq:MLS}) can be expressed as
\begin{align}
    &J^{S_a;L,S}\left( m'_a;p'_{b},m'_b,p'_{c},m'_c \right)\notag\\
    &=F^{[S_a,m'_a]} \left( T_L(p'_b,p'_c,m'_L),\chi_{S}^{\rm tot}(m'_S) \right)\notag\\
    &={q^{'\nu_1}\cdots q^{'\nu_L}\epsilon^*_{\nu_1\cdots\nu_L}(m'_L)}F^{[S_a,m'_a]} \left( \chi_L(m'_L),\chi_{S}^{\rm tot}(m'_S)  \right)B_L(Q_{abc})\notag\\
    &={q^{'\nu_1}\cdots q^{'\nu_L}\epsilon^*_{\nu_1\cdots\nu_L}(m'_L)}C_{Lm'_LSm'_S}^{S_am'_a}C_{S_bm'_bS_cm'_c}^{Sm'_S} \chi_{S_a}(m'_a) B_L(Q_{abc}),   \label{eq:LS}
\end{align}
where $m'_{L}$ is the $z$-component of the orbital angular momentum in $a^*$-frame. Finally, we have 
\begin{align}
\mathcal M_{a\to bc}(m'_a;p'_b,m'_b,p'_c,m'_c) = \sum_{L,S}  A^{S_a}_{L,S}C_{Lm'_LSm'_S}^{S_am'_a}C_{S_bm'_bS_cm'_c}^{Sm'_S}{{\cal Y}_{L,m_L'}(q')}B_L(Q_{abc}),\label{eq:M1to2}
\end{align}
{where
\begin{equation}
    {\cal Y}_{L,m}(q)=q^{\nu_1}\cdots q^{\nu_L}\epsilon^*_{\nu_1\cdots\nu_L}(m)
\end{equation}
is the covariant form of spherical harmonics.}
\section{Sequential decays}\label{sec:IV}
Consider a sequential decay process
\begin{align}
a^{S_a}(p_a,m_a)&\to b^{S_b}(p_b,m_b)\overbrace{+}^{L,S}c^{S_c}(p_c,m_c),\\
c^{S_c}(p_c,m_c)&\to d^{S_d}(p_d,m_d)\overbrace{+}^{L_1,S_1}e^{S_e}(p_e,m_e),
\end{align}
where $p_i$ and $m_i$ are again defined in the laboratory frame and $L_1$ and $S_1$ are the orbital angular momentum and total spin of $d,e$ system.   For later convenience, we first define some other frames based on the pre-defined laboratory frame by Lorentz transformations. At first, we are at a pre-defined laboratory frame, then by a inversed STDB which depends on particle $a$'s momentum $p_a$ in the laboratory frame, we travel to one of the STDF of particle $a$, say $a*$-frame. In this frame particle $a$ is at its STDM $k_a$. Next, by a inversed STDB which depends on particle $c$'s momentum $p_c'$ in $a^*$-frame, we travel to one of the STDF of particle $c$, say $a^*c^*$-frame. In this frame, particle $c$ is at its STDM $k_c$. Finally by a inversed STDB which depends on particle $i$'s momentum $p_i''$ in $a^*c^*$-frame, with $i=d$ or $e$, we travel to one of the STDF of the final state particle $i$, say $a^*c^*i^*$-frame. In this frame, particle $i$ is at its STDM $k_i$. It is worth to mention that all these frames are fixed once the laboratory frame is given since all the Lorentz transformations are the STDBs defined in Eq.(\ref{eq:standboost}) without any ambiguities.

The amplitude can be constructed using the $1\to 2$ decay amplitudes in last section, which reads
\begin{align}
    &\mathcal{M}(m_a'; p'_b,m'_b,p''_d,m''_d,p''_e,m''_e)\notag\\
    &=\sum_{m'_c}\mathcal M_{a\to bc}(m'_a;p'_b,m'_b,p'_c,m'_c) \mathcal M_{c\to de}(m''_c=m'_c;p''_d,m''_d,p''_e,m''_e)f(p_c),\label{eq:seq1}
\end{align}
where $p''_{i\mu}=(\tilde\Lambda^{-1}_{k_c\to p_c'}\tilde\Lambda^{-1}_{k_a\to p_a})_{\mu\nu} p^\nu_i$ and $m_i''(i=c,d,e)$ is the polarization of particle $i$ in the $a^*c^*$-frame, which is obtained from the laboratory frame by sequential Lorentz transformations $\tilde\Lambda^{-1}_{k_i\to p_i''} \tilde\Lambda^{-1}_{k_c\to p_c'}\tilde\Lambda^{-1}_{k_a\to p_a}$.
${\cal M}_{a\to bc}$ and ${\cal M}_{c\to de}$, which are two Lorentz and dirac scalars, are amplitudes defined in Eq.(\ref{eq:M1to2}). Note that $m''_c=m'_c$ since they are both defined in the $a^*c^*$-frame. We have introduced a factor $f(p_c)$ as the propagator of resonance $c$, which can be parameterized in the Breit-Wigner form, Flatt\`e form~\cite{Flatte:1976xu}, Gounatis-Sakurai form~\cite{Gounaris:1968mw} or any other reasonable forms. 

If the process only has one decay chain, Eq.(\ref{eq:seq1}) is enough since averaging the initial states and summing over the final states in terms of $m_i',m_i''$ is equivalent to that in terms of $m_i$. However, if more than one decay chains are possible, we have to transform the amplitude expressed in Eq.(\ref{eq:seq1}) to the one evaluated in a common frame, i.e., in terms of $m_i$, so that the amplitudes for different decay chains can be added coherently. In this way the amplitude reads
\begin{align}
    \mathcal M^{\rm lab}(p_a,m_a;p_b,m_b,p_d,m_d,p_e,m_e)=\sum_{m_b',m_d'',m_e''}&D^{S_b}_{m_b,m_b'}(\Lambda^{a}_{k_b\to k_b})D^{S_d}_{m_d,m''_d}(\Lambda^{ac}_{k_d\to k_d})D^{S_e}_{m_e,m''_e}(\Lambda^{ac}_{k_e\to k_e})\notag\\
    &\times\mathcal{M}(m_a';p'_b,m'_b,p''_d,m''_d,p''_e,m''_e),\label{eq:Mseq}
\end{align}
with $\Lambda^{a}_{k_i\to k_i}=\tilde\Lambda^{-1}_{k_i\to p_i}\tilde\Lambda_{k_a\to p_a}\tilde\Lambda_{k_i\to p'_i}$ and $\Lambda^{ac}_{k_i\to k_i}=\tilde\Lambda^{-1}_{k_i\to p_i}\tilde\Lambda_{k_a\to p_a}\tilde\Lambda_{k_c\to p_c'}\tilde\Lambda_{k_i\to p''_i}$, which can be described using Thomas angle in Ref~\cite{2013srgf.book.....G}.
Note that the amplitude in Eq.(\ref{eq:Mseq}) is expressed in terms of the momenta and polarizations in laboratory frame and hence it can be used directly in the cases where there are different decay chains and interference. The generalization to more decay chains is straightforward. 

\section{Examples}\label{sec:V}
In this section, we give several examples to illustrate how the method we proposed above works.
\subsection{$\phi\to K^+K^-$}
Since the final particles are spinless, the CG coefficients in Eq.(\ref{eq:M1to2}) are all equal to 1 and the amplitude simply reads
\begin{align}
    \mathcal{M}_(m'_a;p'_b,m'_b=0,p'_c,m'_c=0)=q^{\prime\,\nu_1}\epsilon^*_{\nu_1}(m'_a)B_1(Q_{abc}),
\end{align}
which is exactly the same as that in covariant $L$-$S$ scheme. 
\subsection{$J/\psi\to\phi\eta$ or $\gamma\eta$}
Let us label $J/\psi,\phi(\gamma),\eta$ as particle $a,b,c$, respectively. In this case, $(L,S)=(1,1)$, and thus from Eq.(\ref{eq:M1to2}), the amplitude in the $a^*$-frame reads
\begin{align}
     \mathcal M_{J/\psi\to \phi\eta}(m'_a;p'_b,m'_b,p'_c,m'_c=0) &= A^{1}_{1,1}C_{1(m'_a-m'_b)1m'_b}^{1m'_a}{q^{\prime\,\nu_1}\epsilon^*_{\nu_1}(m'_a-m'_b)}B_1(Q_{abc}),\\
     \mathcal M_{J/\psi\to \gamma\eta}(m'_a;p'_b,m'_b,p'_c,m'_c=0) &= A^{1}_{1,1}\sum_{m_b''} {D_{m_b',m_b''}(R(\hat{\bm p}_b'))}C_{1(m'_a-m''_b)1m''_b}^{1m'_a}\notag\\
     &~~~~~~~~~~~~~~~~\times{q^{\prime\,\nu_1}\epsilon^*_{\nu_1}(m'_a-m''_b)}B_1(Q_{abc}).\label{eq:psitogammaeta}
\end{align}

Summing over the initial and final spins we obtain the angular distribution. If the $J/\psi$ is produced in the $e^+e^-$ annihilation, its polarization can only takes the value $\pm1$ and in turn 
\begin{align}
    &\sum_{m_a'=\pm1}\sum_{m_b'=0,\pm1}| \mathcal M_{J/\psi\to \phi\eta}(m'_a;p'_b,m'_b,p'_c,m'_c=0)|^2\notag\\
    =&\sum_{m_a'=\pm1}\sum_{m_b'=\pm1}| \mathcal M_{J/\psi\to \gamma\eta}(m'_a;p'_b,m'_b,p'_c,m'_c=0)|^2\propto 1+\cos^2\theta,
\end{align}
If the $J/\psi$ has equal possibilities for three polarizations, there should be no special direction and thus the angular distribution is expected to be flat, which is verified by the following explicit calculations,
\begin{align}
     &\sum_{m_a'=0,\pm1}\sum_{m_b'=0,\pm1}| \mathcal M_{J/\psi\to \phi\eta_c}(m'_a;p'_b,m'_b,p'_c,m'_c=0)|^2\notag\\
    =&\sum_{m_a'=0,\pm1}\sum_{m_b'=\pm1}| \mathcal M_{J/\psi\to \gamma\eta_c}(m'_a;p'_b,m'_b,p'_c,m'_c=0)|^2\propto1,
\end{align}
One can easily check that they agree with the distributions using the amplitudes constructed in Ref~\cite{Zou:2002yy}.

\subsection{$J/\psi\to b_1\eta$}
Let us label $J/\psi,b_1,\eta$ as particle $a,b,c$, respectively. In this case, $(L,S)=(0,1)$ and $(2,1)$, by the constraint of parity. The amplitudes are
\begin{align}
    \mathcal{M}_{0,1}^1&=A_{0,1}^1C^{1m_a'}_{1m_b'0m_c'}\\
    \mathcal{M}_{2,1}^1&=A_{2,1}^1C^{1m_a'}_{2m_L'1m_S'}C^{1m_S'}_{1m_b'0m_c'}q^{\prime\nu_1}q^{\prime\nu_2}\epsilon^*_{\nu_1\nu_2}(m_L')B_2(Q_{abc}),
\end{align}

 \subsection{$J/\psi\to\gamma\eta_c\to\gamma\Lambda\bar\Lambda$}
Let us label $J/\psi,\gamma,\eta_c,\Lambda,\bar\Lambda$ as particle $a,b,c,d,e$, respectively. For $J/\psi\to\gamma\eta_c$, the amplitude is the same as Eq.(\ref{eq:psitogammaeta}). For $\eta_c\to\Lambda\bar\Lambda$, we have $(L,S)=(0,0)$ and the amplitude in the $a^*c^*$-frame reads
\begin{align}
    \mathcal M_{\eta_c\to \Lambda\bar\Lambda}(m''_c;p''_d,m''_d,p''_e,m''_e)=  A^{0}_{0,0}C_{\frac12 m''_d\frac12 m''_e}^{0,0},
\end{align}
being flat as expected. Then the amplitude for the sequential decay from Eq.(\ref{eq:seq1}) is
\begin{align}
    \mathcal{M}(m_a';p'_b,m'_b,p''_d,m''_d,p''_e,m''_e)=A\,\sum_{m_b^{\prime\dagger}} {D^{1}_{m_b',m_b^{\prime\dagger}}(R(\hat{\bm p}_b'))} C_{\frac12 m''_d\frac12 m''_e}^{0,0}C_{1(m'_a-m_b^{\prime\dagger})1m_b^{\prime\dagger}}^{1m'_a}\notag\\{q^{\prime\,\nu_1}\epsilon^*_{\nu_1}(m'_a-m_b^{\prime\dagger})}B_1(Q_{abc})f(p_c),\label{eq:ampexm1}
\end{align}
Since this process has only one decay chain, summing over initial and final polarizations can be performed directly to Eq.(\ref{eq:ampexm1}).

 \subsection{$J/\psi\to\gamma X(2^{-+}) \to\gamma\Lambda\bar\Lambda$ }

Let us first focus on the $J/\psi\to\gamma X(2^{-+})\to\gamma\Lambda\bar\Lambda$ and label $J/\psi,\gamma,X(2^{-+}),\Lambda,\bar\Lambda$ as particle $a,b,c,d,e$, respectively. Here $X(2^{-+})$ is a imaginary meson-like state, assumed for discussion. For $J/\psi\to\gamma X(2^{-+})$, we have $(L,S)=(1,1),(1,2),(3,2),(3,3)$ and the amplitude in the $a^*$-frame reads
\begin{align}
    \mathcal{M}_{1,1}^1 &= A_{1,1}^1 \sum_{m_b^{\prime\dagger}} {D^{1}_{m_b',m_b^{\prime\dagger}}(R(\hat{\bm p}_b'))} C^{1m_a'}_{1m_L'1m_S'}C^{1m_S'}_{1m_b^{\prime\dagger}2m_c'} q^{\prime\,\nu_1}\epsilon^*_{\nu_1}(m_L')B_1(Q_{abc})\\
    \mathcal{M}_{1,2}^1 &= A_{1,2}^1 \sum_{m_b^{\prime\dagger}} {D^{1}_{m_b',m_b^{\prime\dagger}}(R(\hat{\bm p}_b'))} C^{1m_a'}_{1m_L'2m_S'}C^{2m_S'}_{1m_b^{\prime\dagger}2m_c'} q^{\prime\,\nu_1}\epsilon^*_{\nu_1}(m_L')B_1(Q_{abc})\\
    \mathcal{M}_{3,2}^1 &= A_{3,2}^1 \sum_{m_b^{\prime\dagger}} {D^{1}_{m_b',m_b^{\prime\dagger}}(R(\hat{\bm p}_b'))} C^{1m_a'}_{3m_L'2m_S'}C^{2m_S'}_{1m_b^{\prime\dagger}2m_c'} q^{\prime\,\nu_1}q^{\prime\,\nu_2}q^{\prime\,\nu_3}\epsilon^*_{\nu_1\nu_2\nu_3}(m_L')B_3(Q_{abc})\\
    \mathcal{M}_{3,3}^1 &= A_{3,3}^1 \sum_{m_b^{\prime\dagger}} {D^{1}_{m_b',m_b^{\prime\dagger}}(R(\hat{\bm p}_b'))} C^{1m_a'}_{3m_L'3m_S'}C^{3m_S'}_{1m_b^{\prime\dagger}2m_c'} q^{\prime\,\nu_1}q^{\prime\,\nu_2}q^{\prime\,\nu_3}\epsilon^*_{\nu_1\nu_2\nu_3}(m_L')B_3(Q_{abc}).
\end{align}
Note that the master decay amplitude of $J/\psi\to\gamma X$ reads $\mathcal{M}_{a \to bc}(m_a'; p_b',m_b',p_c',m_c')=\sum_{LS} \mathcal{M}^{1}_{L,S}$. The decay of $X(2^{-+})\to\Lambda\bar\Lambda$ has $(L,S)=(2,0),(2,1)$ and the amplitude in the $a^*c^*$-frame reads
\begin{align}
    \mathcal{M}_{2,0}^2 &= A_{2,0}^2 C^{2m_c''}_{2m_L''0m_S''}C^{0m_S'}_{\frac{1}{2}m_d''\frac{1}{2}m_e''} q^{\prime\prime\,\nu_1}q^{\prime\prime\,\nu_2}\epsilon^*_{\nu_1\nu_2}(m_L'')B_2(Q_{cde})\\
    \mathcal{M}_{2,1}^2 &= A_{2,1}^2 C^{2m_c''}_{2m_L''1m_S''}C^{1m_S'}_{\frac{1}{2}m_d''\frac{1}{2}m_e''} q^{\prime\prime\,\nu_1}q^{\prime\prime\,\nu_2}\epsilon^*_{\nu_1\nu_2}(m_L'')B_2(Q_{cde}),
\end{align}
and we note the master decay amplitude of $X(2^{-+})\to\Lambda\bar\Lambda$ as $\mathcal{M}_{c \to de}(m_c''=m_c';p_d'',m_d'',p_e'',m_e'')=\sum_{LS} \mathcal{M}^{2}_{L,S}$.
Since the amplitude should be a Lorentz invariant complex number, it allows us to evaluate amplitude of different levels in different frames and multiply them directly to form total amplitude like Eq.(\ref{eq:seq1}). The master amplitude for $J/\psi\to\gamma X(2^{-+}) \to\gamma\Lambda\bar\Lambda$ reads
\begin{align}
    &\mathcal{M}
    (m_a'; p'_b,m'_b,p''_d,m''_d,p''_e,m''_e)\notag\\
    &=\sum_{m'_c}\mathcal M_{a\to bc}(m'_a;p'_b,m'_b,p'_c,m'_c) \mathcal M_{c\to de}(m''_c=m'_c;p''_e,m''_e,p''_e,m''_e)f(p_c),
\end{align}
where $f(p_c)$ is the propagator for $X(2^{-+})$ state. If we would like to add this process coherently with other decay chain, for example, $J/\psi \to \Lambda^{*}\bar\Lambda\to\gamma\Lambda\bar\Lambda$, we need to align the polarizations of the final particles to common STDFs according to Eq.(\ref{eq:Mseq}), in which all final polarizations are aligned in the STDFs which connect the laboratory frame by STDB
\begin{align}
    \mathcal M^{\rm lab}(p_a,m_a;p_b,m_b,p_d,m_d,p_e,m_e)=\sum_{m_b',m_d'',m_e''}&D^{1}_{m_b,m_b'}(\Lambda^{a}_{k_b\to k_b})D^{\frac{1}{2}}_{m_d,m''_d}(\Lambda^{ac}_{k_d\to k_d})D^{\frac{1}{2}}_{m_e,m''_e}(\Lambda^{ac}_{k_e\to k_e})\notag\\
    &\times\mathcal{M}(m_a';p'_b,m'_b,p''_d,m''_d,p''_e,m''_e),
\end{align}
with $\Lambda^{a}_{k_i\to k_i}=\tilde\Lambda^{-1}_{k_i\to p_i}\tilde\Lambda_{k_a\to p_a}\tilde\Lambda_{k_i\to p'_i}$ and $\Lambda^{ac}_{k_i\to k_i}=\tilde\Lambda^{-1}_{k_i\to p_i}\tilde\Lambda_{k_a\to p_a}\tilde\Lambda_{k_c\to p_c'}\tilde\Lambda_{k_i\to p''_i}$. Note that $p'_b,p''_d,p''_e$ can be expressed in terms of $p_a,p_b,p_d,p_e$.

\subsection{$\psi(2S)\to\Omega\bar\Omega$}
In this case, $(L,S)=(0,1),(2,1),(2,3)$ and $(4,3)$. The amplitudes are expressed as
\begin{align}
    \mathcal{M}_{0,1}^1&=A_{0,1}^1C^{1m_a'}_{\frac{3}{2}m_b'\frac{3}{2}m_c'}\\
    \mathcal{M}_{2,1}^1&=A_{2,1}^1C^{1m_a'}_{2m_L'1m_S'}C^{1m_S'}_{\frac{3}{2}m_b'\frac{3}{2}m_c'}q^{\prime\nu_1}q^{\prime\nu_2}\epsilon^*_{\nu_1\nu_2}(m_L')B_2(Q_{abc})\\
    \mathcal{M}_{2,3}^1&=A_{2,1}^1C^{1m_a'}_{2m_L'3m_S'}C^{3m_S'}_{\frac{3}{2}m_b'\frac{3}{2}m_c'}q^{\prime\nu_1}q^{\prime\nu_2}\epsilon^*_{\nu_1\nu_2}(m_L')B_2(Q_{abc})\\
    \mathcal{M}_{4,3}^1&=A_{4,3}^1C^{1m_a'}_{4m_L'3m_S'}C^{3m_S'}_{\frac{3}{2}m_b'\frac{3}{2}m_c'}q^{\prime\nu_1}q^{\prime\nu_2}q^{\prime\nu_3}q^{\prime\nu_4}\epsilon^*_{\nu_1\nu_2\nu_3\nu_4}(m_L')B_4(Q_{abc}),
\end{align}
Summing over all possible polarizations will of course yield flat angular distribution and if only $m_a'=\pm$ are considered, we have
\begin{align}
    \sum_{m_a'=\pm1}\sum_{m_b',m_c'=-3/2}^{3/2}| \mathcal M_{0,1}^1|^2&\propto1,\\
    \sum_{m_a'=\pm1}\sum_{m_b',m_c'=-3/2}^{3/2}| \mathcal M_{2,1}^1|^2&\propto5-3\cos^2\theta,\\
    \sum_{m_a'=\pm1}\sum_{m_b',m_c'=-3/2}^{3/2}| \mathcal M_{2,3}^1|^2&\propto5-\cos^2\theta,\\
    \sum_{m_a'=\pm1}\sum_{m_b',m_c'=-3/2}^{3/2}| \mathcal M_{4,3}^1|^2&\propto11-5\cos^2\theta,
\end{align}
Let us emphasis again that the spin wavefunctions of $\Omega$ and $\bar\Omega$ are not necessary here, contrary to the covariant $L$-$S$ scheme~\cite{Zou:2002yy} where the explicit expressions for spin coupling is needed. If fermions with higher spins are included, such explicit expressions will be much complicated, difficult to construct and in these cases, the advantages of our method are revealed.

\subsection{$\Lambda_b \to P^+_c(\frac{1}{2}^{-}) K^- \to J/\psi p K^-$}
Similarly with the case of $J/\psi\to\gamma X(2^{-+}) \to\gamma\Lambda\bar\Lambda$, let us first label $\Lambda_b,P^+_c(\frac{1}{2}^{-}),K^-,J/\psi,p$ as particle $a,b,c,d,e$, respectively. Here $P^+_c(\frac{1}{2}^{-})$ is a possible pentaquark state with $J^P=\frac{1}{2}^-$. For $\Lambda_b \to P^+_c(\frac{1}{2}^{-}) K^-$, we have $(L,S)=(0,\frac{1}{2})$ and $(1,\frac{1}{2})$, and the amplitude in the $a^*$-frame reads
\begin{align}
    \mathcal{M}_{0,\frac{1}{2}}^1 &= A_{0,\frac{1}{2}}^1 C^{\frac{1}{2}m_a'}_{0m_L'\frac{1}{2}m_S'}C^{\frac{1}{2}m_S'}_{\frac{1}{2}m_b^{\prime}0m_c'}B_0(Q_{abc}),\\
    \mathcal{M}_{1,\frac{1}{2}}^1 &= A_{1,\frac{1}{2}}^1 C^{\frac{1}{2}m_a'}_{1m_L'\frac{1}{2}m_S'}C^{\frac{1}{2}m_S'}_{\frac{1}{2}m_b^{\prime}0m_c'} q^{\prime\,\nu_1}\epsilon^*_{\nu_1}(m_L')B_1(Q_{abc}).
\end{align}
The master decay amplitude of $\Lambda_b \to P^+_c(\frac{1}{2}^{-}) K^-$ reads $\mathcal{M}_{a \to bc}(m_a'; p_b',m_b',p_c',m_c')=\sum_{LS} \mathcal{M}^{1}_{L,S}$. The decay of $P^+_c(\frac{1}{2}^{-})\to J/\psi p$ has $(L,S)=(1,\frac{1}{2})$ and $(1,\frac{3}{2})$, and the amplitude in the $a^*b^*$-frame reads
\begin{align}
    \mathcal{M}_{1,\frac{1}{2}}^2 &= A_{1,\frac{1}{2}}^2 C^{\frac{1}{2}m_c''}_{1m_L''\frac{1}{2}m_S''}C^{\frac{1}{2}m_S'}_{1m_d''\frac{1}{2}m_e''} q^{\prime\prime\,\nu_1}\epsilon^*_{\nu_1}(m_L'')B_1(Q_{cde}),\\
    \mathcal{M}_{1,\frac{3}{2}}^2 &= A_{1,\frac{3}{2}}^2 C^{\frac{1}{2}m_c''}_{1m_L''\frac{3}{2}m_S''}C^{\frac{3}{2}m_S'}_{1m_d''\frac{1}{2}m_e''} q^{\prime\prime\,\nu_1}\epsilon^*_{\nu_1}(m_L'')B_1(Q_{cde}).
\end{align}
The master decay amplitude of $P^+_c(\frac{1}{2}^{-})\to J/\psi p$ reads $\mathcal{M}_{b \to de}(m_b''=m_b';p_d'',m_d'',p_e'',m_e'')=\sum_{LS} \mathcal{M}^{2}_{L,S}$. The total amplitude for $\Lambda_b \to P^+_c(\frac{1}{2}^{-}) K^- \to J/\psi p K^-$ reads exactly the same as Eq.(\ref{eq:seq1}),
\begin{align}
    &\mathcal{M}
    (m_a'; p'_c,m'_c,p''_d,m''_d,p''_e,m''_e)\notag\\
    &=\sum_{m'_b}\mathcal M_{a\to bc}(m'_a;p'_b,m'_b,p'_c,m'_c) \mathcal M_{b\to de}(m''_b=m'_b;p''_e,m''_e,p''_e,m''_e)f(p_b),
\end{align}
where $f(p_b)$ is the propagator for $P^+_c(\frac{1}{2}^{-})$ state. If we would like to add this process coherently with other decay chains, say, $\Lambda_b \to J/\psi \Lambda^{*} \to J/\psi p K^-$, alignments according to Eq.(\ref{eq:Mseq}) need to be done,
\begin{align}
    \mathcal M^{\rm lab}(p_a,m_a;p_c,m_c,p_d,m_d,p_e,m_e)=\sum_{m_c',m_d'',m_e''}&D^{0}_{m_c,m_c'}(\Lambda^{a}_{k_c\to k_c})D^{1}_{m_d,m''_d}(\Lambda^{ab}_{k_d\to k_d})D^{\frac{1}{2}}_{m_e,m''_e}(\Lambda^{ab}_{k_e\to k_e})\notag\\
    &\times\mathcal{M}(m_a';p'_c,m'_c,p''_d,m''_d,p''_e,m''_e),
\end{align}
with $\Lambda^{a}_{k_i\to k_i}=\tilde\Lambda^{-1}_{k_i\to p_i}\tilde\Lambda_{k_a\to p_a}\tilde\Lambda_{k_i\to p'_i}$ and $\Lambda^{ab}_{k_i\to k_i}=\tilde\Lambda^{-1}_{k_i\to p_i}\tilde\Lambda_{k_a\to p_a}\tilde\Lambda_{k_b\to p_b'}\tilde\Lambda_{k_i\to p''_i}$. One can check the consistency with the Dalitz-plot decomposition\cite{JPAC:2019ufm}. 

\subsection{$B \to J/\psi K \pi \pi$}
Here we just use $B \to X(3872)(1^{++}) K \to J/\psi \rho K \to J/\psi \pi_1 \pi_2 K$ as example. Other chains can be constructed in a similar way. First we label $B,X(3872),K,J/\psi,\rho,\pi_1,\pi_2$ as particle $a,b,c,d,e,f,g$, respectively. For $B \to X(3872) K$, we have $(L,S)=(1,1)$. The partial amplitude $\mathcal{M}^{1}_{L,S}$ in the $a^*$-frame could be constructed using Eq.(\ref{eq:M1to2}).
And the master decay amplitude is noted as $\mathcal{M}_{a \to bc}(m_a'; p_b',m_b',p_c',m_c')=\sum_{LS} \mathcal{M}^{1}_{L,S}$. Similarly, The decay of $X(3872)\to J/\psi \rho$ has $(L,S)=(0,1),(2,1)$ and $(2,2)$, and the partial amplitude $\mathcal{M}^{2}_{L,S}$ in the $a^*b^*$-frame could be constructed. The master decay amplitude is noted as $\mathcal{M}_{b \to de}(m_b''=m_b''; p_d'',m_d'',p_e'',m_e'')=\sum_{LS} \mathcal{M}^{2}_{L,S}$. The decay of $\rho \to \pi \pi$ has $(L,S)=(1,0)$, and master amplitude in $a^*b^*e^*$-frame is constructed with partial amplitude as $\mathcal{M}_{e \to fg}(m_e'''=m_e''; p_f''',m_f''',p_g''',m_g''')=\sum_{LS} \mathcal{M}^{3}_{L,S}$. Finally, the total amplitude for $B \to X(3872)(1^{++}) K \to J/\psi \rho K \to J/\psi \pi_1 \pi_2 K$ reads
\begin{align}
    &\mathcal{M}
    (m_a'; p'_c,m'_c,p''_d,m''_d,p'''_f,m'''_f,p'''_g,m'''_g)=\sum_{m'_b,m''_e}\mathcal M_{a\to bc}(m'_a;p'_b,m'_b,p'_c,m'_c)\notag\\
    & \times \mathcal M_{b\to de}(m''_b=m'_b;p''_d,m''_d,p''_e,m''_e)\mathcal M_{e\to fg}(m'''_e=m''_e;p'''_f,m'''_f,p'''_g,m'''_g)f(p_b)f(p_e)
\end{align}
where $f(p_b)$ and $f(p_e)$ is the propagator for $X(3872)$ and $\rho$, respectively. The alignment is done by
\begin{align}
    \mathcal M^{\rm lab}(p_a,m_a;p_c,m_c,p_d,m_d,p_f,m_f,p_g,m_g)=\sum_{m_c',m_d'',m_f''',m_g'''}D^{0}_{m_c,m_c'}(\Lambda^{a}_{k_c\to k_c})D^{1}_{m_d,m''_d}(\Lambda^{ab}_{k_d\to k_d})\notag\\
    \times D^{0}_{m_f,m'''_f}(\Lambda^{abe}_{k_f\to k_f})D^{0}_{m_g,m'''_g}(\Lambda^{abe}_{k_g\to k_g})
    \mathcal{M}(m_a';p'_c,m'_c,p''_d,m''_d,p'''_f,m'''_f,p'''_g,m'''_g),
\end{align}
with matching transformations $\Lambda^{a}_{k_i\to k_i}=\tilde\Lambda^{-1}_{k_i\to p_i}\tilde\Lambda_{k_a\to p_a}\tilde\Lambda_{k_i\to p'_i}$, $\Lambda^{ab}_{k_i\to k_i}=\tilde\Lambda^{-1}_{k_i\to p_i}\tilde\Lambda_{k_a\to p_a}\tilde\Lambda_{k_b\to p_b'}\tilde\Lambda_{k_i\to p''_i}$ and $\Lambda^{abe}_{k_i\to k_i}=\tilde\Lambda^{-1}_{k_i\to p_i}\tilde\Lambda_{k_a\to p_a}\tilde\Lambda_{k_b\to p_b'}\tilde\Lambda_{k_e\to p''_e}\tilde\Lambda_{k_i\to p'''_i}$. 

\subsection{$D_s^+\to\phi e^+\nu_e$}
The proposed method can also be applied to leptonic and semi-leptonic decays. Let us take the semi-leptonic process $D_s^+\to\phi e^+\nu_e$ via intermediate $W^+$ exchange as example where $D_s^+,\phi,W^+,e^+,\nu_e$ are denoted by particle $a,b,c,d,f$, respectively. For the first vertex, $D_s^+\to \phi W^+$, $(L,S)=(0,0),(1,1)$ or $(2,2)$ and the corresponding amplitudes read
\begin{align}
    \mathcal{M}_{0,0}^0&=A_{0,0}^0C^{0,0}_{1m_b'1m_c'},\\
    \mathcal{M}_{1,1}^0&=A_{2,1}^0C^{0,0}_{1m_L'1m_S'}C^{1m_S'}_{1m_b'1m_c'}q^{\prime\nu_1}\epsilon^*_{\nu_1}(m_L')B_1(Q_{abc}),\\
    \mathcal{M}_{2,2}^1&=A_{2,1}^0C^{0,0}_{2m_L'2m_S'}C^{2m_S'}_{1m_b'1m_c'}q^{\prime\nu_1}q^{\prime\nu_2}\epsilon^*_{\nu_1\nu_2}(m_L')B_2(Q_{abc}),
\end{align}
The second vertex $W^+\to e^+\nu_e$ can be precisely described by the electro-weak theory and the amplitudes reads
\begin{align}
    \mathcal M(p_c'',m_c'';p_d'',m_d'',p_f'',m_f'')=\frac{e}{\sqrt{2}\sin\theta_W}\chi_1^\mu(p_c'',m_c)\bar\chi_{1/2}(p_f'',m_f)\gamma_\mu\frac{1-\gamma^5}{2}\chi_{1/2}(p_d'',m_d).
\end{align}
{Note that the amplitude for the above weak interaction can also be constructed using the method we proposed above but we need not to do that since we have the explicit and accurate expression.}

\section{Summary}\label{sec:VI}
The covariant $L$-$S$ scheme to construct the decay amplitudes is very convenient and useful for extracting the resonance parameters from the high-quality data. In this scheme, the amplitudes with different total spin and orbital angular momentum in the final states are constructed separately and orthogonal to each other. In this work, we overcome some drawbacks in the pioneer works~\cite{Zou:2002ar,Zou:2002yy} and proposed a modified version of the amplitude construction. Our method holds the Lorentz invariance, avoids the complication of the explicit spin couplings, and if only massive particles involved, separates the orbital and spin angular momenta thoroughly.

We also would like to emphasis that the proposed amplitude construction method only depends on the masses and spins of involved particles and has nothing to do with the detailed dynamics. Therefore, it will also be useful for leptonic or semi-leptonic decays.

We are grateful to the fruitful discussions with Liao-Yuan Dong, Feng-Kun Guo, Bai-Cian Ke, Hai-Bo Li, Jia-Jun Wu and Bing-Song Zou. This work is supported by the NSFC and the Deutsche Forschungsgemeinschaft (DFG, German Research Foundation) through the funds provided to the Sino German Collaborative Research Center TRR110 Symmetries
and the Emergence of Structure in QCD (NSFC Grant No.12070131001, DFG Project-ID 196253076-TRR 110), by the
NSFC Grants Nos. 11835015, 11875054, 11935018, 12047503, 12192263, by
the Chinese Academy of Sciences (CAS) under Grant No.XDB34030000, by Joint Large-Scale Scientific Facility Fund of the NSFC and CAS under Grant No.U2032104, and by China Postdoctoral Science Foundation under Grant No. 119103S408.

\bibliography{ref}
\end{document}